\documentclass[journal]{IEEEtranTIE}
\usepackage[utf8]{inputenc}
\usepackage[T1]{fontenc}
\usepackage{graphicx}
\usepackage{cite}
\usepackage{picinpar}
\usepackage{amsmath}
\usepackage{url}
\usepackage{flushend}
\usepackage{colortbl}
\usepackage{soul}
\usepackage{multirow}
\usepackage{pifont}
\usepackage{color}
\usepackage{alltt}
\usepackage[hidelinks]{hyperref}
\usepackage{enumerate}
\usepackage{siunitx}
\usepackage{breakurl}
\usepackage{epstopdf}
\usepackage{pbox}
\usepackage{amsthm}
\usepackage{algorithm}
\usepackage{url}
\usepackage{multirow}
\usepackage{amssymb}
\usepackage{subfig}
\usepackage{algorithm}
\usepackage[noend]{algpseudocode}
\usepackage{amsfonts}
\usepackage{filecontents}
\usepackage{mathtools}
\usepackage{xspace}

\begin{document}

\title{Analysis of Anomalous Behavior in Network Systems Using Deep Reinforcement Learning with CNN Architecture}

\author{
	\vskip 1em
	
Mohammad Hossein Modirrousta, Parisa Forghani Arani, and Mahdi Aliyari Shoorehdeli, \emph{Senior Member}, \emph{IEEE}

		\thanks{Mohammad Hossein Modirrousta (e-mail: mohammadbc@email.kntu.ac.ir) is with the Faculty of Electrical Engineering, K. N. Toosi University of Technology, Tehran, Iran.
  
          Parisa Forghani Arani (e-mail: pfor0005@student.monash.edu) is with the Faculty of Information Technology, Monash University, VIC 3800, Australia.		
		
		Mahdi Aliyari Shoorehdeli (e-mail: aliyari@kntu.ac.ir) is with the Faculty of Mechatronics Engineering, K. N. Toosi University of Technology, Tehran, Iran.}}

\maketitle
	
\begin{abstract}
        In order to gain access to networks, different types of intrusion attacks have been designed and improved. Computer networks have become increasingly important in daily life due to the increasing reliance on them. In light of this, it is quite evident that algorithms with high detection accuracy and reliability are needed for various attack types. The purpose of this paper is to develop an intrusion detection system that is based on deep reinforcement learning. Based on the Markov decision process, the proposed system can generate informative representations suitable for classification tasks based on vast data. This paper inspects reinforcement learning from two perspectives: deep Q learning and double deep Q learning. Different experiments have demonstrated that the proposed systems have an accuracy of $99.17\%$ over the UNSW-NB15 dataset in both approaches, an improvement over previous methods based on contrastive learning and LSTM-Autoencoders. The performance of the model trained on UNSW-NB15 has also been evaluated on BoT-IoT dataset, resulting in competitive performance.
\end{abstract}



\definecolor{limegreen}{rgb}{0.2, 0.8, 0.2}
\definecolor{forestgreen}{rgb}{0.13, 0.55, 0.13}
\definecolor{greenhtml}{rgb}{0.0, 0.5, 0.0}

\section{Introduction}

    \label{sec:into}
    \IEEEPARstart{I}{n} view of the importance of the telecommunication services deployed in data networks, proper and reliable operation is crucial regarding economic and social impacts. Since this situation and cybersecurity issues have a significant impact, it is imperative to develop new algorithms that can detect intrusions and are also efficient in terms of resources, considering the new nature of massive and complex networks, including the Internet of Things \cite{bhuyan2013network}.

In the cyber domain, there is an increasing number of advanced attackers who pose threats, requiring new Intrusion Detection Systems (IDS) methods that have automated and intelligent network intrusion detection strategies to handle them. Therefore, there is an increased demand for intelligent agent-based intrusion detection and prevention systems capable of evolving and improving without involving humans \cite{hou2020low}, \cite{brundage2018malicious}.

Intrusion detection systems can be classified from different aspects, for instance, based on their deployment \cite{Attention-basedRL}, the place they are set up, or whether they are active or passive \cite{Attention-basedRL}. IDS' operation method can also categorize them into signature- and anomaly-based systems. A signature-based IDS detects destructive code based on predefined patterns called signatures. This method effectively performs static detection with a lower false-positive rate (FPR) \cite{context-awareRL}. The downsides of these systems are their need for manual updating of the signature database and their inability to detect unexpected or zero-day attacks \cite{ids_category}, \cite{buczak2015survey}. Alternatively, the anomaly-based system detects destructive behavior based on deviations from standard functioning. These systems can detect zero-day attacks. Despite their advantages, anomaly-based systems have the disadvantage of having difficulty detecting traffic accurately. Consequently, their false-positive rate is usually high \cite{context-awareRL}.

In recent years, many researchers applied traditional machine learning algorithms to intrusion detection to increase efficiency and reduce the false-positive and false-negative rates. The drawbacks of these methods are that they need to provide better performance for large datasets and multiple classifications, as they require high-dimensional representations \cite{context-awareRL}.

As science and technology advanced, deep learning (DL) techniques became a solution to the mentioned shortcomings and, therefore, a popular solution for designing IDSs. These methods can generally be categorized into supervised and unsupervised learning. Supervised learning usually lacks in detecting unknown attacks \cite{LSTM}.
Unsupervised learning allows learning feature representations without labels. Thus, they have a better performance on zero-day attacks. However, these models can only be generalized to some datasets \cite{choi2019unsupervised}.

Reinforcement Learning (RL) can be an answer to mentioned limits. This branch of machine learning learns from and adapts to new changes efficiently and automatically and finds optimal behavior using reward-based sequential decision-making. This way, RL-based IDS has no difficulty detecting new attacks \cite{context-awareRL}. There are various algorithms for RL, namely Q-Learning \cite{QLearning} and SARSA \cite{SARSA}. However, the traditional RL cannot build complex security models as it lacks scalability \cite{Attention-basedRL}. On the bright side, converging RL with deep learning resulting in deep reinforcement learning (DRL) brings out a new chapter that can handle complicated logic and models. A DRL-based IDS can identify remarkably advanced cyberattacks \cite{Attention-basedRL}.

This paper uses a DRL model to empower the intrusion detection system. The model has been tested with different policies, deep Q-learning (DQL) and double deep Q-learning (DDQL), which the comparison is available in section~\ref{sec:sim}. The proposed network is first trained on the UNSW-15 dataset. Then, with the help of transfer learning, the network is tested on the BoT-IoT dataset. This study aims to show that the trained model can be used on other datasets containing various features with the lowest computational cost. Wherefore, such a network can be used in online frameworks.

\section{Related Work}
    \label{sec:related}
    As the need for a safe IoT connection strikes, researchers have paid more and more attention to this field. Much research has been done involving artificial intelligence and machine learning in the field of intrusion detection. In the following, some of the latest and most relevant papers to this research has highlighted. The publications have been classified according to their point of commonality to our research.

\textbf{Usage of UNSW-NB-15 Dataset:} In \cite{LSTM}, the UNSW-NB15 dataset demonstrates the external network traffic cyber-attacks, and the car hacking dataset illustrates the in-vehicle communications. After pre-processing and feature reduction, \cite{LSTM} leverages an LSTM layer to capture the latent temporal and spatial non-linearities in the sequence of feature vectors. The results show that \cite{LSTM} proposed structure works perfectly for both types of cyber-attacks. On the other hand, while the article has claimed to identify a wide range of cyber-attacks, it has not depicted its results class-wise.

\textbf{Usage of BoT-IoT Dataset:} In \cite{botiot1}, a multi-class feed-forward neural network classifier is used to identify four categories of attacks: DoS, DDoS, data gathering, and data theft. Also, through a transfer learning-based approach, the encoding of high-dimensional categorical features is extracted. \cite{botiot1} has used the BoT-IoT dataset for validating their model. Their results show that their suggested architecture can only partially classify some attack classes.

In \cite{botiot2} paper, a three-stage model is built to anomaly detect intrusion in IoT networks. A CNN-based multi-class classification model uses 1D, 2D, and 3D convolutional neural networks to identify 15 attack types from normal traffic data. \cite{botiot2} validated their work using the BoT-IoT, IoT Network Intrusion, MQTT-IoT-IDS2020, and IoT-23 datasets. They also leverage transfer learning to implement binary and multi-class classification with a pre-trained CNN multi-class model. Their results show high accuracy, precision, recall, and F1 scores for the binary and multi-class models.

\textbf{Usage of Transfer Learning:} A recently published paper suggests a CNN-based architecture alongside transfer learning. \cite{transferCNN} converts each chunk of the time-based vehicle network traffic data into images and assigns the most repeated label as the block's label. They use VGG16, VGG19, Xception, Inception, and InceptionResnet, which have been pre-trained on the ImageNet dataset as their base. Although their reported results are high, converting chunks of tabular data into images and their labeling technique carries some data loss.

\textbf{Usage of Q-Learning:} \cite{qtransfer} suggest a transfer learning Q-learning-based structure. They adopted deep transfer learning to lighten the dataset insufficiency problem. Although the headlines of \cite{qtransfer} look similar to our paper, they have only focused on DDoS attacks while we cover a full range of attacks in IoT network intrusions.

Another newly issued article \cite{pooraccDQL} combines Q-learning-based reinforcement learning with a deep feed-forward neural network. They claim their model is equipped with an ongoing auto-learning capability. \cite{pooraccDQL} results are based on the NSL-KDD dataset. However, their obtained accuracy must fulfill the expectations with 0.81 and 0.88 for the Normal and R2L classes, respectively. Similarly, their confusion matrix reports a noticeable amount of miss-classed data.

\textbf{Summary:} The above review of current literature shows that reinforcement learning and Q-learning approaches have not received their deserved attention in implementing intrusion detection solutions for IoT networks.
In this direction, this paper's transfer learning-based deep Q-learning model is proposed. This advanced robust IDS works anomaly-based and intends to achieve high accuracy and low FPR in detecting a full range of network intrusions. One of our main achievements is maintaining our results on different datasets containing various attack types described with various features through transfer learning. The outcome of our designed model has been compared with numerous state-of-art models, and its superiority is shown.

\section{Background}
    \label{sec:back}

\subsection{Reinforcement Learning} \label{ssec:RL}
Reinforcement learning (RL) is a reward-based approach that relies on the interaction between an agent and a given environment, intending to maximize the numerical reward \cite{sutton2018reinforcement}. As a result of feedback from the environment, the agent learns its behavior and subsequently attempts to improve its actions. The solution to reinforcement learning problems is to work out a policy (i.e., mapping from state to action) that maximizes the accumulation of rewards in the long run. There are five relevant entities in a reinforcement learning problem: state, action, reward, policy, and value. Generally, a reinforcement learning problem is modeled by a Markov decision process.

\subsubsection{Q-Learning} \label{sssec:QL}
In reinforcement learning, the Q-learning algorithm is used in which agents learn policies as they transition between states \cite{watkins1992q}. By exploring all possible feasible actions related to the different states of the agent, we can determine the optimal set of policies. This algorithm maintains its value continuously by updating a Q-value according to the next state of the algorithm and the greedy action. Q-functions essentially accept a variety of arguments, including state vectors (${s}$), action vectors (${a}$), reward vectors (${r}$), and learning rates (${\gamma}$). The discount factor is then calculated for the Q-value. Due to the requirement of high dimensionality, however, Q-learning-based systems perform poorly in large state spaces.

\subsubsection{Deep Q Network (DQN)}
\label{sssec:DQN}
There is a technique known as Deep Q-Network (DQN) that can be used to deal with this problem, and it is a networked Q-learning algorithm that combines reinforcement learning with a class of artificial neural networks known as Deep Q Networks \cite{mnih2015human}.

\begin{figure}[ht]
  \includegraphics[clip, trim=2cm 17cm 0.5cm 2.4cm,width=0.5\textwidth,height=4.5cm, scale=1.5]{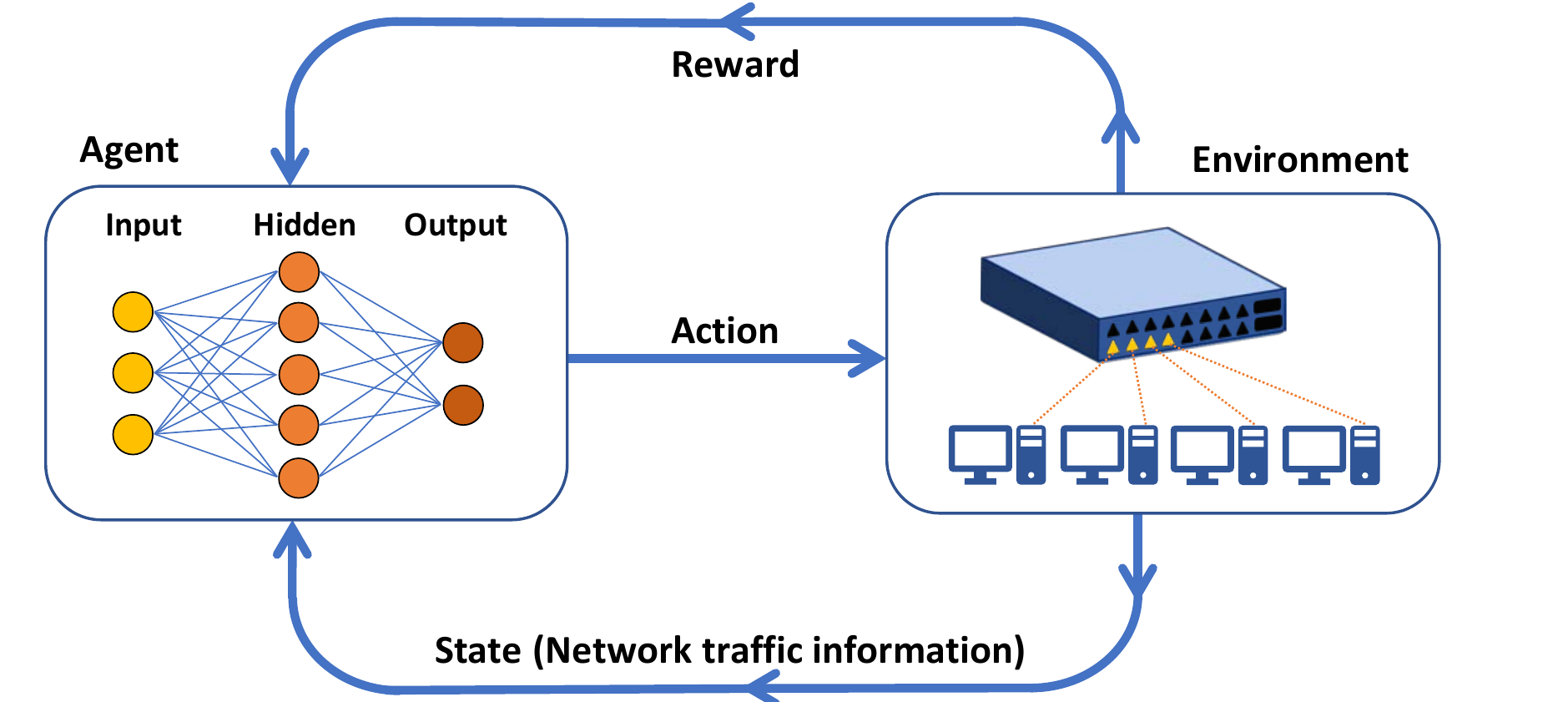}
  \caption{An overview of the process flow of deep Q-networks.}\label{fig:Agent}
\end{figure}

DQN's process flow is shown in Figure \ref{fig:Agent}, where input is received from the environment as a state, which is then used for calculating what action to take based on the weights and then shared with the environment. The environment grants a positive reward if the action is favorable, or it is penalized with a negative reward if the action is not favorable. The reward is also used to update the weights of the DNN to ensure its performance is improved.

A key objective of the DQN algorithm is to choose actions that will result in the maximum cumulative future reward possible. Using the DNN, we can approximate the optimal action-value function ${Q^{*}(s, a)}$. Using the variables shown in Figure~\ref{fig:Agent}, the optimal action-value function can be calculated as follows:

\begin{equation}
\resizebox{.9\hsize}{!}{$Q^{*}(s, a)=\max _{\pi} \mathbb{E} \underbrace{\left[r_{t}+\gamma r_{t+1}+\gamma^{2} r_{t+2}+\cdots\right.}_{\mathrm{R}} \mid s_{t}=s, a_{t}=a, \pi]$}
\end{equation}

A theory states that the maximum sum of rewards ${r_{t}}$ discounted by ${\gamma}$, which could be achieved by taking action ${a}$ from state ${s}$, is possible if the policy is considered equal to $\pi=P(a \mid s)$. In the above equation, ${\mathbb{E}}$ represents the expectation of the random variable ${R}$, and $P(a \mid s)$ represents the conditional probability of action ${a}$ in a given state ${s}$.

The DQN algorithm has achieved impressive performance in practice when playing Atari games. This algorithm primarily integrates DNNs with Q-Learning algorithms to build self-aware decision policies ${\pi}$ that map state ${s}$ to an action ${a}$ in such a way that ${a=\pi(s)}$ \cite{mnih2015human}, \cite{mnih2013playing}.

Achieving maximum future reward, ${R_{t}}$, at time ${t}$ is the objective of the adaptive method, which is discounted by a factor of ${\gamma}$ at each step and is defined as follows:  
\begin{equation}
R_{t}=\sum_{t^{\prime}=t}^{T} \gamma^{t^{\prime}-t} r_{t^{\prime}}
\end{equation}
In the above equation, ${T}$ represents the end of the time step and ${\gamma}$ represents the discount factor. Generally, the less value of ${\gamma}$, the more important the agent places on the current reward. 

According to the Q-learning algorithm, the action-value function ${Q^{\pi}(s, a)}$ is as follows:  

\begin{equation}
Q^{\pi}(s, a)=\mathbb{E}_{\pi}\left[R_{t} \mid s_{t}=s, a_{t}=a ; \pi\right]
\end{equation}
Under the policy ${\pi}$, ${\mathbb{E}_{\pi}}$ represents the expected return. 

A Q-learning algorithm's temporal-difference updating is expressed as follows:  
\begin{equation}
Q_{j+1}(s, a)=\mathbb{E}_{\pi}\left[r+\gamma \max _{a^{\prime}} Q_{j}\left(s^{\prime}, a^{\prime}\right) \mid s, a\right]
\end{equation}

Specifically, ${Q(s, a)}$ indicates the action-value function.

According to the Bellman equation, the optimal action-value function ${Q^{*}(s, a) = max_{\pi}Q^{\pi}(s, a)}$ is characterized by  
\begin{equation}
    Q^{*}(s, a)=\mathbb{E}_{\pi}\left[r+\gamma \max _{a^{\prime}} Q^{*}\left(s^{\prime}, a^{\prime}\right) \mid s, a\right]
\end{equation}
To obtain the action-value function  ${Q^{\pi}(s, a),}$ in the policy ${\pi}$, neural networks are used in DQN.  
\begin{equation}
    Q(s, a ; \boldsymbol{\theta}) \approx Q^{\pi}(s, a)
\end{equation}
In this case, ${\theta}$ represents the parameters of the neural network model.

As a part of the process of the agent's interaction with the environment, the $\epsilon$-greedy algorithm is used to make sure the agent maintains a balance between exploration (random action ${a}$) and exploitation (action ${a}$ with $\operatorname{arg}_{a} \max \left(Q\left(s_{t},\{a\}\right)\right)$) and generates a series of experiences as it interacts with the environment \cite{sutton2018reinforcement}. The experience samples are stored in a database. A random selection of them is then made for updating the DNN model. 

The mean-square error (MSE) is used as an objective function when updating the model's parameters. 
\begin{equation}
L_{j}\left(\boldsymbol{\theta}_{j}\right)=\mathbb{E}\left[\left(y_{j}-Q\left(s, a ; \boldsymbol{\theta}_{j}\right)\right)^{2}\right]
\end{equation}
Assume that ${y}$ is the target ${Q}$ value function and that it can be calculated as follows:   
\begin{equation}
y_{j}=r+\gamma \max _{a^{\prime}} Q\left(s^{\prime}, a^{\prime} ; \boldsymbol{\theta}_{j-1}\right)
\end{equation}

The gradient descent algorithm is used to update the objective function to determine the optimal policy ${\pi^{*} = \operatorname{arg}_{a} \max ( Q^{*}(s, a,\theta))}$. In this regard, the gradient updating is calculated as follows: 
\begin{equation}
\nabla_{\boldsymbol{\theta}_{j}} L\left(\boldsymbol{\theta}_{j}\right)=\mathbb{E}_{\sim U}\left[\left(y-Q\left(s, a ; \boldsymbol{\theta}_{j}\right)\right) \nabla_{\boldsymbol{\theta}_{j}} Q\left(s, a ; \boldsymbol{\theta}_{j}\right)\right]
\end{equation}
In this case, the random uniform sampling is represented by ${\sim U}$.



\subsubsection{Double Deep Q Network (DDQN)} \label{sssec:DDQN}
In many aspects, DDQN is comparable to DQN. DDQN differs from DQN in that it employs two neural networks: one implements a current Q function, and the other implements a target Q function. In other words, the target Q function is a copy of the current Q function with a delayed timing loop made after a certain number of training runs. With the target Q function, gradient descent over objective functions can be performed without recursive dependence on training networks, avoiding the moving target effect \cite{van2016deep}.
 
\subsection{Using Markov Decision Processes for classification } \label{ssec:cmdp}
In computer gaming, control, recommendation systems, and communication, deep reinforcement learning (DRL) has achieved tremendous success by combining the potential of DL with the potential of reinforcement learning (RL) \cite{luong2019applications}, \cite{vu2015power}, \cite{precup2020evolving}, \cite{liang2019deep}. Nevertheless, the DRL solution is rarely used for classification tasks since it deals with the problem of sequential decision-making. The classification Markov decision process (CMDP) has been developed and defines a standard classification problem as a sequential decision-making problem. The MLP model trained in this framework performed better than typical MLP models trained by backpropagation \cite{wiering2011reinforcement}. As a result, the DRL applications automatically adapt their decision-making policies to the environment, so the cumulative future reward can be maximized without adding new features while improving their generalization capability.    

Assuming the CMDP framework with the tuple $\{S, A, P, R\}$, we can consider an intrusion detection problem as a sequential decision-making problem, where ${S}$, ${A}$, ${P}$, and ${R}$ represent the state space, action space, transition probability, and reward function, respectively.

\textbf{State space ${S}$:} It is determined by samples of training. If an episode begins, the agent will be given its initial state  $s_{1} \in S$ based on the sample ${x_{1}}$. In the same way, environment state, ${s_{t}}$, represents sample ${x_{t}}$. The environment shuffles the order of samples when a new episode begins so that ${T}$ samples are collected to create the training data sequence. 

\textbf{Action space ${A}$:} It has a list of all categories of recognized actions, and the action $a_{t} \in A$ equates with an attack label or a normal label. Accordingly, ${A}$ consists of $\{1,2, \ldots, L\}$, whereas ${L}$ represents the overall condition of the intrusion data.

\textbf{Transition probability ${P}$:} A transition model is referred to as $P: S \times A \longrightarrow S$. As far as CMDP is concerned, the transition probability is deterministic. Once the agent has obtained the current state $s_{t} \in S$ and taken the appropriate action, the agent will receive the next state $s_{t+1} \in S$ according to the order of samples in the mini-batch. All samples in the mini-batch are then processed sequentially until the problem is resolved.

\textbf{Reward function ${R}$:} Evaluating the outcome of the agent's actions in the reward function promotes the agent's learning of the policy ${\pi}$ and enables it to make appropriate decisions in given states. As soon as the label of the state is correctly detected, this work results in a positive reward from the environment; otherwise, a negative reward is obtained. As a result, the reward function is designed in the following manner: 

\begin{equation}
R\left(s_{t}, a_{t}, l_{t}\right)= \begin{cases}1, & \text { if } a_{t}=l_{t} \\ -1, & \text { if } a_{t} \neq l_{t}\end{cases}
\end{equation}

$l_{t}$ indicates an attack or normal label.

According to the preceding definitions, intrusion detection can be described as a sequential decision-making problem in CMDP. The goal is to maximize accumulated rewards by choosing the most optimal policy. The solution to this problem can be achieved using DRL.

\subsection{Transfer learning}
\label{ssec:Transferlearning}
Transfer learning, a machine learning technique, modifies the starting point of the model's training to another pre-trained model \cite{Des&Develop}. In this manner, the model does not start from scratch, resulting in an investment of time and resources \cite{Idrissi2021-ui}. Transfer Learning has proved helpful in convolutional neural networks (CNN). As a CNN model gets deeper with each layer, the patterns learned by the inner layers become more general. Thus, the weights learned by the bottom layers can be transferred to other models with diverse tasks. Transfer Learning can be further successful with fine-tuning. Fine-tuning allows the model to learn the higher-order features by unfreezing only a few top layers of the model. Therefore, the pre-trained model will fit the target task or dataset better \cite{transferCNN}.

\subsection{Dataset}
\label{ssec:dataset}

\subsubsection{UNSW-NB15 Dataset}
\label{sssec:unsw}
In the Australian Cyber Security Centre (ACCS), the IXIA PerfectStorm tool created the UNSW-NB15 dataset in 2015. This dataset has been designed based on a comprehensive network environment for generating attack activities \cite{UNSW}.

Data are presented under two categories, attack and normal. The attack category further breaks down into nine subcategories. The data flow is described in 49 features, 47 attack-related. Table~\ref{tab:unsw_size} provides a detailed description of the data type and size of the UNSW-NB15 dataset.

\begin{table} [ht]
	\caption{Sizes and data type of UNSW-NB15 dataset}
	\renewcommand{\arraystretch}{1.5}
	\label{tab:unsw_size}
	\centering
	\resizebox{0.3\textwidth}{!}{%
		\setlength\tabcolsep{1pt}
		\begin{tabular}{cccccccccc}
			 \hline
			Category&Number of Packets \\\hline
			Normal & 677785  \\
			Fuzzers & 5051  \\
			 Dos & 1167  \\
			 Exploits & 5408 \\
			 Generic & 7522 \\
			 Reconnaissance & 1759 \\
			\hline
			Total&698692\\
			\hline

		\end{tabular}
	}
\end{table}

\subsubsection{BoT-IoT Dataset}
\label{sssec:botiot}
In the Cyber Range Lab of The center of UNSW Canberra Cyber, Koroniotis et al. used the Node-red tool to simulate the network behavior of IoT devices and proposed the BoT-IoT dataset in 2018. This dataset incorporates legitimate and simulated IoT network traffic in five IoT scenarios. \cite{BoT-IoT}

Data are presented under two categories, attack and normal. The attack category further breaks down into four subcategories. The data flow is described in 49 features, 47 attack-related. Table~\ref{tab:botiot_size} provides a detailed description of the data type and size of the BoT-IoT dataset.

\begin{table} [ht]
	\caption{Sizes and data type of BoT-IoT dataset}
	\renewcommand{\arraystretch}{1.5}
	\label{tab:botiot_size}
	\centering
	\resizebox{0.3\textwidth}{!}{%
		\setlength\tabcolsep{1pt}
		\begin{tabular}{cccccccccc}
			 \hline
			Category&Number of Packets \\\hline
			DDoS & 1926624  \\
			DoS & 1650260  \\
			 Normal & 477 \\
			 Theft & 79 \\
			 Reconnaissance & 91082 \\
			\hline
			Total&3668522\\
			\hline
		\end{tabular}
	}
\end{table}

\section{Proposed method}
    \label{sec:method}
\begin{figure*}
  \includegraphics[clip, trim=0cm 12cm 0cm 0.5cm,width=\textwidth, scale=5]{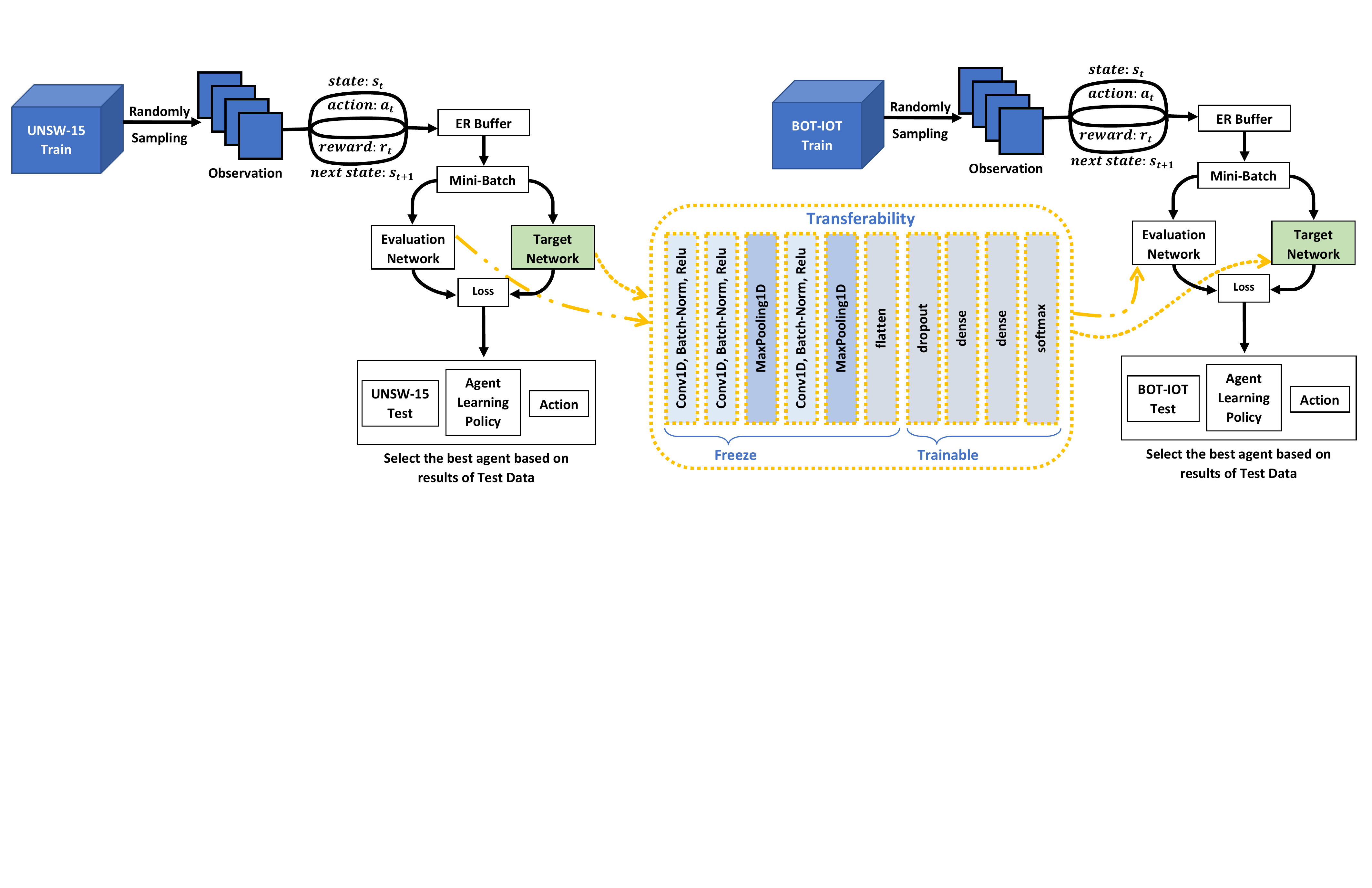}
  \caption{Overall illustration of proposed DQN and DDQN where only the DDQN contains the "Target Network" (in green).}\label{fig:frm}
\end{figure*}

The previous section examined reinforcement learning and deep reinforcement learning techniques, as well as DQN and DDQN algorithms, to classify our desired data. Our next step will be to classify the following data using the model we obtained from the algorithm's convergence. An overview of our framework is shown in Figure \ref{fig:frm}. Below are the specific steps involved in the RL intrusion detection model:  
\begin{enumerate}
\item In the preprocessing of the UNSW-NB15 intrusion detection dataset, one-hot encoding, normalization of the data, converting IP features to numerical features, and some other processes are performed. 

\item RL Model parameters are initialized, and the neural network model structure is set up.

\item Configuring the environment in CMDP form and setting the parameters of the CMDP format.

\item Iterations are set so that the current network in DQN and the current network and target network in DDQN will converge at the end of the iterations.

\item Inputting the feature representation to the classification layer and getting the results for the UNSW-NB15 dataset.

\item Select the most appropriate loss function based on the performance of Mean Square Error, KL Divergence, Huber, and CategoricalCrossentropy.

\item Get the feature extraction model and save the weights.

\item To perform transfer learning on the other dataset, use the feature extraction model and the appropriate classification layer.

\end{enumerate}

Algorithm \ref{alg:cap} describes a pseudocode of the algorithms.

\begin{algorithm}
\caption{Proposed DQN and DDQN}\label{alg:cap}
\begin{algorithmic}[1]
\Require
        \State States $\mathcal{S} = \{1, \dots, s_n\}$
        \State Actions $\mathcal{A} = \{1, \dots, a_n\}, A: \mathcal{S} \Rightarrow \mathcal{A}$
        \State Reward function $R: \mathcal{S} \times \mathcal{A} \rightarrow \mathbb{R}$
        \State Probabilistic transition function $P: \mathcal{S} \times \mathcal{A} \rightarrow \mathcal{S}$
        \State Labels  $\mathcal{L} = \{l_1, \dots, l_n\}$ 
        \State Learning rate $\alpha \in [0, 1]$
        \State Discounting factor $\gamma \in [0, 1]$
        \State ER buffer $B$, Experience Replay Memory $M$
\State Current model $Q_{\theta}$
\State Target model $Q_{\theta^{\prime}}$
        \Procedure{Q Learning} {$\mathcal{S}$, $A$, $R$, $P$}
\State \textbf{input:} $X$=$\left\{\left(s_1, l_1\right),\left(s_2, l_2\right) \ldots \ldots \ldots\left(s_n, l_n\right)\right\}$
\For {each iteration}
\State Shuffle $X$

\For {sampled minibatch $\{s_k\}_{k=1}^{N}$}

\State Calculate $\pi$ based on Q and 
\Statex \hspace{11mm} exploration strategy ($\pi(s) \gets Q(s, a)$)
\State $a_{t} \gets \pi(s_{t})$ 
\State $r_{t} \gets R(s_{t}, a_{t})$ \Comment{Receive the reward}
\State $s_{t+1} \gets P(s_{t}, a_{t})$ \Comment{Receive the new state}   \State Store $(s_t,a_t,r_t,s_{t+1})$ in $B$
\State Random sample $(s_i,a_i,r_i,s_{i+1})$ from $M$
\If{Use DQN}
\State Compute $Q$ value:
\Statex \hspace{11mm} $Q^{*}(s_{i}, a_{i})=r_{i}+\gamma \max _{a^{\prime}} Q_{{\theta}}\left(s_{i+1}, a_{i+1}\right)$
\State $L\left(\boldsymbol{\theta}\right)=(Q^*\left(s_i, a_i\right)-Q_\theta\left(s_i, a_i\right))^2$      
\State Perform gradient descent on $L\left(\boldsymbol{\theta}\right)$ 
\State Update network parameters
\EndIf
\If{Use DDQN}
\State Compute target $Q$ value:
\Statex \hspace{11mm} $
{Q^*}\left( {{s_i},{a_i}} \right) = (1 - \alpha )\,{Q_\theta }\left( {{s_i},{a_i}} \right)\, $
\Statex \hspace{11mm} $+ \alpha \left( {{r_i} 
 + \gamma {Q_\theta }\left( {{s_{i + 1}},{{{\mathop{\rm argmax}\nolimits} }_{{a^\prime }}}{Q_{{\theta ^\prime }}}\left( {{s_{i + 1}},{a^\prime }} \right)} \right)} \right)$ 
\State $L\left(\boldsymbol{\theta}\right)=(Q^*\left(s_i, a_i\right)-Q_\theta\left(s_i, a_i\right))^2$      
\State Perform gradient descent on $L\left(\boldsymbol{\theta}\right)$ 
\State Update current network parameters
\State Update target network parameters
\EndIf
\EndFor
\State Save network(s) parameters ${\theta}$
\State Use ${\theta}$ for evaluate on new data
\EndFor
 \EndProcedure
\end{algorithmic}
\end{algorithm}





\section{Simulations}
    \label{sec:sim}
    \subsection{Preprocessing}
The raw data form in almost every experimental dataset does not match the input form of the neural network in question. Therefore, preprocessing is required.  

\paragraph{One-hot encoding}
The labels of our dataset include string names such as Normal, DDOS, and others, which are unsuitable for ML algorithms and need to be interpreted numerically. The critical factor in this conversion is not prioritizing any label over others which will happen if we assign a number to each category. Here, One Hot Encoding shines through. One Hot Encoding creates a table where the number of columns equals the number of unique labels, and the number of rows equals the number of data points in the dataset. Then, in each row, it only sets the cell value of the corresponding label to one.

\paragraph{Rectifying incorrect cell values}
Another common mistake in datasets is when a label prints out incorrectly. The UNSW-NB15 dataset needed various error-catching in this field. For instance, the \textit{ct-ftp-cmd} column with the expected outputs of integers from 0 to 8 printed out null instead of 7. Similarly, the \textit{attack-cat} column printed null data instead of the label Normal. In another case, the \textit{ct-flw-http-mthd} column reported more than 1.3 million null data points. Since deducting this many data points is not advisable, we have replaced the null values with the mean of all other column values.

\paragraph{Removing incorrect data points or column}
Every dataset has data points with misleading or wrong values that should be detected and removed. In the UNSW-NB15 dataset, we have deleted any data points with non-numerical \textit{S-port} or \textit{D-port} values. These very few data points contained hexadecimal numbers and represented normal network traffic. They got removed as they did not follow the rest of the dataset's nature and made other features of the dataset lose their meaning. Other problematic data points are missing values (NaN-values) for any dataset's features. These data points got detected and removed. On another occasion, we removed the \textit{is-ftp-login} column as it contained more than 1.4 million null data points. Since we opted to use one-hot encoding, and this column indicates whether an FTP session has been used, deducting it from the dataset would not be problematic.

\paragraph{Normalization}
Normalization allows us to process the data more effectively and faster. One of the most straightforward normalization techniques is scaling the minimum and maximum of the dataset to a specific number, usually 0 and 1. This linear transformation keeps all data relationships and makes the optimization problem more numerically stable.

\subsection{Training}
For both DQN and DDQN algorithms, we set the batch size to 256 and epochs to 25 in the CMDP framework. Furthermore, we initialized $\epsilon$ to 0.8 and $\gamma$ to 0.001. In order to train the Adam optimization algorithm, we use a learning rate of 0.001 as part of the training process. For DDQN, we set the initial value of $\alpha$ to 1 with a decay rate of 0.99. In Table \ref{tab:str}, we describe the neural networks' parameters that share the same structure as the current DQN network and DDQN's current and target networks. The ReLU activation is used to activate hidden neurons. At each execution step, $80\%$ of the data is used for training, while the remainder is used for testing.

\begin{table} [!htb]
	\caption{The structure of neural networks in DQN and DDQN}\label{tab:str}
	\renewcommand{\arraystretch}{1.5}
	\label{tab:reg}
	\centering
	\resizebox{0.48\textwidth}{!}{%
		\setlength\tabcolsep{3pt}
		\begin{tabular}{cccccccccc}
		     \hline
			Layers&UNSW-NB15 pack&BoT-IoT pack\\\hline
			Conv1D & 16@(1×2) & \textquotedbl\\
			Conv1D& 32@(1×2)& \textquotedbl\\
			Max Pool1D& 1×2 &\textquotedbl\\
			Conv1D& 64@(1×2)&\textquotedbl \\
			Max Pool1D& 1×2 &\textquotedbl \\
			Flatten\\
			Dense& 40&5 \\
			Dense& \_\_&5 \\

			\hline
			Dense (softmax) & 6 &5 \\
			 \hline

		\end{tabular}
	}
\end{table}

\subsection{Loss functions}
In order to estimate the Q value, we have used different loss functions for the neural network. For this reason, we compare the performance and find the best cost function that provides the best results. The training conditions of the network and the other hyperparameters have been kept consistent for a fair comparison. Our purpose here is to explain the function of the selected costs.
\paragraph{Mean Squared Error (MSE)}
One of the most straightforward yet commonly used loss functions is MSE which averages the square of the difference between the model's prediction and the ground truth. This loss function is efficient in learning outlier data.
\begin{equation}
MSE=\frac{1}{N}\sum_{i=1}^{N} \left(y_{i}-\hat{y}_{i}\right)^2
\end{equation}
Where \textit{N} is the number of test samples, \textit{y} is the ground truth, and \textit{\^{y}} is the predicted value.

\paragraph{Categorical Cross-Entropy}
Softmax Loss or Categorical Cross-Entropy is the combination of Softmax activation and Cross-Entropy loss. This loss function is mainly used for multi-classed problems with one-hot labels where each data point belongs to one class only.
\begin{equation}
CE=-\sum_{i=1}^{S} y_{i}\cdot\log\hat{y}_{i}
\end{equation}
Where \textit{S} is the number of scalar values in the model output, \textit{y} is the ground truth, and \textit{\^{y}} is the predicted value.

\paragraph{Kullback–Leibler divergence}
KL divergence calculates the information loss when approximating one distribution to another. KL divergence is not symmetric and thus is not a distance metric.
\begin{equation}
D_{KL}\left(p||q\right)=\sum_{i=1}^{N} p(x_i)\log{\left(\frac{p(x_i)}{q(x_i)}\right)}
\end{equation}
Where \textit{N} is the number of test samples, \textit{q(x)} is the true distribution, and \textit{p(x)} is the approximate distribution.

\paragraph{Huber}
Huber loss uses Mean Absolute Error (MAE) for greater values of $\lvert y-f(x)\rvert$ loss and MSE for otherwise. By this means, the model is not as sensitive to larger losses while containing a quadratic function for the smaller loss values. Huber loss stands between MSE and MAE.
\begin{equation}
L_{\delta}\left(y, f(x)\right)= \begin{cases}{\frac{1}{2}}\left(y-f(x)\right)^2 & \text { for } \lvert{y-f(x)}\rvert\leq\delta, \\ \delta\lvert{y-f(x)}\rvert-{\frac{1}{2}}\delta^2 & \text { otherwise. } \end{cases}
\end{equation}
Where \textit{$\delta$} is the penalty value, \textit{y} is the ground truth, and \textit{f(x)} is the predicted value.

According to Table \ref{tab:loss1}, although the Huber and KLDivergence cost functions performed better in the precision criterion in the DQN framework, the categorical cross-entropy cost function performed better in the remaining evaluation criteria. In most applications, the MSE cost function is sufficient, but in our application, its performance is weaker than that of the other modes. A similar problem was encountered with the MSE cost function in the DDQN framework. Similar to the DQN framework, the categorical cross-entropy cost function also improved network training and performance on test data, as shown in Table \ref{tab:loss2}.
\begin{table} [!htb]
	\caption{Comparison of DQN's performance with different loss functions}\label{tab:loss1}
	\renewcommand{\arraystretch}{1.5}
	\label{tab:reg}
	\centering
	\resizebox{0.48\textwidth}{!}{%
		\setlength\tabcolsep{3pt}
		\begin{tabular}{cccccccccc}
			\hline
			method & Accuracy& Precision & Recall& F1 Score \\ \hline
			MSE  &0.9870&0.9936&0.9870&0.9903\\
			CategoricalCrossentropy  &\textbf{0.9917}&{0.9946}&\textbf{0.9917}&\textbf{0.9932}\\
			KLDivergence &0.9894&\textbf{0.9950}&0.9894&0.9922\\
		   Huber  &{0.9912}&\textbf{0.9950}&{0.9912}&0.9931\\
			\hline		
			
		\end{tabular}
	}
\end{table}

\begin{table} [!htb]
	\caption{Comparison of DDQN's performance with different loss functions}\label{tab:loss2}
	\renewcommand{\arraystretch}{1.5}
	\label{tab:reg}
	\centering
	\resizebox{0.48\textwidth}{!}{%
		\setlength\tabcolsep{3pt}
		\begin{tabular}{cccccccccc}
			\hline
			method & Accuracy& Precision & Recall& F1 Score \\ \hline
			MSE  &0.9739&0.9912&0.9739&0.9825\\
			CategoricalCrossentropy  &\textbf{0.9917}&\textbf{0.9952}&\textbf{0.9917}&\textbf{0.9934}\\
			KLDivergence &0.9736&0.9926&0.9736&0.9830\\
		   Huber  &0.9908&0.9948&0.9908&0.9927\\
			\hline		
			
		\end{tabular}
	}
\end{table}

\subsection{Comparison to state-of-the-art methods}

We tested different cost functions in the preceding section to obtain the best results. These results are collected here in order to compare them with similar studies. Details of the data can be found in Table \ref{tab:unsw_size}. 

As a result, our DQN and DDQN models perform better than the other methods in most cases. The accuracy, recall, and F1 score improved by approximately $3\%$, $2\%$, and $1\%$ more than the LSTM-Autoencoder model. We have also found that our model has less than a $1\%$ improvement in all evaluation criteria over the contrastive model . This trend can also be observed with the DDQN model. Furthermore, DDQN outperforms DQN in precision and F1 score by an insignificant margin. This improvement is believed to result from using the reinforcement learning environment and the reward mechanism for processing each mini-batch in the CMDP format. As a side note, even though we bear more computational load in DDQN due to the interaction between two neural networks, the performance difference is not statistically significant compared to DQN.

\begin{table} [!htb]
	\caption{Comparison of our best proposed methods with other techniques on UNSW-NB15}\label{tab:transfer}
	\renewcommand{\arraystretch}{1.5}
	\label{tab:reg}
	\centering
	\resizebox{0.48\textwidth}{!}{%
		\setlength\tabcolsep{3pt}
		\begin{tabular}{cccccccccc}
			\hline
			Data & method & Accuracy& Precision & Recall& F1 Score \\ \hline
			UNSW-NB15  &  LSTM AutoEncoder \cite{ashraf2020novel} &0.96&\textbf{1.00}&0.97&0.98\\
			& Contrastive learning \cite{lotfi2022network}& { 0.9886}&{ 0.9919}&{0.9886}&{0.9900}\\
			& Our DQN model &\textbf{0.9917}&{0.9946}&\textbf{0.9917}&{0.9932}\\
			
	   	   & Our DDQN model &\textbf{0.9917}&\textbf{0.9952}&\textbf{0.9917}&\textbf{0.9934}\\\\
			\hline

			\hline			
			
		\end{tabular}
	}
\end{table}
\subsection{Transferability}

Transfer learning offers several advantages, including reducing the number of parameters in the training process and achieving convergence more quickly. With the help of this technique, the final model volume is reduced, and network prediction is accomplished faster in practice. Following our discussion in section \ref{sec:method}, in two approaches, DQN and DDQN, we use the most appropriate model using the best model obtained from different cost functions. It can be seen in Table \ref{tab:str} that the convolutional layers placed before the flattening layer are frozen. We have removed the dense layer with 40 neurons used for the first data and replaced it with two dense layers with five neurons. The implementation is based on BoT-IoT data, which is discussed in section \ref{sec:back}. We have 9,587 parameters in this case, of which only 3,875 can be trained.

As can be seen, reinforcement learning frameworks have outperformed previous works on this dataset, and the evaluation criteria have also increased. Despite its complexity and interaction with the two networks, it is noteworthy that DDQN has yet to provide better performance than DQN. Although DQN has a more straightforward implementation, it is a better choice for this dataset due to its simplicity.

\begin{table} [!htb]
	\caption{Comparison of transferability of designed system to BoT-IoT dataset}\label{tab:transfer}
	\renewcommand{\arraystretch}{1.5}
	\label{tab:reg}
	\centering
	\resizebox{0.48\textwidth}{!}{%
		\setlength\tabcolsep{3pt}
		\begin{tabular}{cccccccccc}
			\hline
			Data & method & Accuracy& Precision & Recall& F1 Score \\ \hline
			
			BoT-IoT & Contrastive learning \cite{lotfi2022network}& {0.9983}&{  0.9983}&{0.9983}&{0.9982}\\
			& TSODE \cite{fatani2021iot} &0.9904&0.9904&0.9904&0.9904 \\
						&Our DQN model &\textbf{0.9996}&\textbf{0.9996}&\textbf{0.9996}&\textbf{0.9996}\\
			&Our DDQN model &{0.9984}&{0.9985}&{0.9984}&{0.9985}\\
			\hline			
			
		\end{tabular}
	}
\end{table}


\section{Conclusions}
    \label{sec:con}
    This paper proposes a deep reinforcement learning method for intrusion detection. We trained a convolutional neural network in classification Markov decision process format using two different approaches: a deep Q network and a double deep Q network. This representation is then fed into a classification head trained on a labeled dataset. 
Using the proposed method, we achieved state-of-the-art results in the multiclass classification task, with an overall accuracy of 0.9917.
Lastly, we demonstrated that the trained CNN could produce efficient, hidden representations for input patterns drawn from different datasets in both DQN and DDQN. The proposed method has the advantage of being transferrable, allowing it to detect new classes of intrusion using other datasets with excellent efficiency, obtaining an accuracy of 0.9996. All in all, this feature makes our network an attractive option for real-world applications.




\bibliographystyle{Bibliography/IEEEtranTIE}
\bibliography{Bibliography/IEEEabrv,Bibliography/BIB_xx-TIE-xxxx}\ 

\end{document}